%% file: conference_101719.tex
\documentclass[10pt,conference]{IEEEtran}
\IEEEoverridecommandlockouts

\newif\ifreviewversion
\reviewversionfalse

\usepackage{cite}
\usepackage{amsmath,amssymb}
\usepackage{graphicx}
\usepackage{booktabs}
\usepackage{array}
\usepackage{multirow}
\usepackage{tabularx}
\usepackage[flushleft]{threeparttable}
\usepackage{float}
\usepackage{stfloats}
\usepackage{xcolor}
\usepackage{tcolorbox}
\tcbuselibrary{skins,breakable}
\usepackage{xurl}
\usepackage[hidelinks]{hyperref}
\ifreviewversion
\usepackage[switch]{lineno}
\fi
\usepackage[shortlabels]{enumitem}
\usepackage{xspace}
\usepackage{comment}
\usepackage{booktabs}
\usepackage{balance}

\setlist[itemize]{leftmargin=1em,itemsep=0.2em,topsep=0.35em,parsep=0pt,partopsep=0pt}

\graphicspath{{figures/}}

\newcommand{\gso}{GSO\xspace}
\newcommand{\sweperf}{SWE-Perf\xspace}
\newcommand{\sweff}{SWE-\allowbreak fficiency\xspace}
\newcommand{\rqone}{RQ1\xspace}
\newcommand{\rqtwo}{RQ2\xspace}
\newcommand{\rqthree}{RQ3\xspace}

\definecolor{orange-web}{RGB}{245,159,0}
\definecolor{sagegreen}{RGB}{132,169,140}
\definecolor{lemonyellow}{RGB}{250,218,94}
\definecolor{skyblue}{RGB}{125,173,224}
\definecolor{coral}{RGB}{231,111,81}
\definecolor{lavender}{RGB}{150,123,182}
\definecolor{mintgreen}{RGB}{129,199,132}
\definecolor{peach}{RGB}{244,177,131}
\definecolor{steelblue}{RGB}{70,130,180}
\definecolor{rosegold}{RGB}{183,110,121}
\colorlet{boxcolor}{sagegreen}  
\newtcolorbox{takeawaybox}[1][]{
  enhanced,
  attach boxed title to top left={xshift=4mm,yshift=-2mm},
  colback=boxcolor!10,
  colframe=boxcolor!60,
  colbacktitle=boxcolor!80,
  coltitle=white,
  fonttitle=\bfseries\small,
  boxed title style={size=small, colframe=boxcolor!80, sharp corners},
  sharp corners,
  boxrule=0.8pt,
  left=4pt, right=4pt, top=7pt, bottom=4pt,
  boxsep=1pt,
  before skip=4pt,
  after skip=4pt,
  breakable,
  title={#1}
}
\newcommand{\findingbox}[2]{%
\begin{takeawaybox}[#1]
#2
\end{takeawaybox}}

\begin{document}

\title{Are Performance-Optimization Benchmarks Reliably Measuring Coding Agents?}

\ifreviewversion
\author{\IEEEauthorblockN{Anonymous Author(s)}}
\else
\author{
\begin{tabular}{@{}>{\centering\arraybackslash}p{0.31\textwidth}
                >{\centering\arraybackslash}p{0.31\textwidth}
                >{\centering\arraybackslash}p{0.31\textwidth}@{}}
Zhi Chen &
Zhensu Sun\textsuperscript{*} &
Yuling Shi \\
Singapore Management University &
Singapore Management University &
Shanghai Jiao Tong University \\
Singapore, Singapore &
Singapore, Singapore &
Shanghai, China \\
zhi.chen.2023@smu.edu.sg &
zssun@smu.edu.sg &
yuling.shi@sjtu.edu.cn
\end{tabular}\\[1.2ex]
\begin{tabular}{@{}>{\centering\arraybackslash}p{0.31\textwidth}
                @{\hspace{0.08\textwidth}}
                >{\centering\arraybackslash}p{0.31\textwidth}@{}}
David Lo &
Lingxiao Jiang \\
Singapore Management University &
Singapore Management University \\
Singapore, Singapore &
Singapore, Singapore \\
davidlo@smu.edu.sg &
lxjiang@smu.edu.sg
\end{tabular}
\thanks{\textsuperscript{*}Corresponding author.}
}
\fi

\maketitle
\pagestyle{plain}
\thispagestyle{plain}
\ifreviewversion
\linenumbers
\fi

\input{sections/00_abstract}

\begin{IEEEkeywords}
software engineering benchmarks, performance optimization, coding agents, benchmark validity
\end{IEEEkeywords}

\input{sections/01_introduction}
\input{sections/03_study_design}
\input{sections/04_rq1}
\input{sections/05_rq2}
\input{sections/06_rq3}
\input{sections/07_discussion}
\input{sections/08_threats}
\input{sections/09_related_work}
\input{sections/10_conclusion}
\input{sections/11_data_availability}

\balance
\bibliographystyle{IEEEtran}
\bibliography{references}

\end{document}

%% file: sections/00_abstract.tex
\begin{abstract}
Repository-level performance-optimization benchmarks such as \sweperf{},
\gso{}, and \sweff{} evaluate coding agents by applying patches to real
repositories and comparing runtime against unoptimized baselines and official
reference patches. Their leaderboard scores are increasingly used as evidence of coding-agent
progress, but those scores can conflate runtime instability, benchmark-specific
scoring rules, and how many tasks are already solved by at least one public
submission. We audit these issues across the three
benchmarks.
First, we replay the official reference patches for 740 code optimization tasks
across four common types of Google Cloud machines. Most benchmark tasks can be
replayed, but their reference patches satisfy the original benchmark validity
rules in every cross-machine replay for only 39/102 \gso{} tasks, 11/140
\sweperf{} tasks, and 411/498 \sweff{} tasks; \sweperf{} is especially fragile
because many reference patches produce close-to-zero runtime changes.
Second, we show that public submission rankings depend strongly on
the benchmark scoring rule. Among eight public submissions shared by \gso{} and
\sweff{}, the official rankings disagree on 9 of 28 pairwise submission
comparisons, and
\sweff{}'s leaderboard scoring rule assigns the worst ten tasks overly high
score weights of 58.5\%--82.8\%.
Third, looking across 10 public submissions for each task, we find that at
least one submission matches or beats the reference patch on 85.3\% (384/450) of
replay-valid \gso{} and \sweff{} tasks, and beats the unoptimized base code on
99.8\% (449/450).
Our study complements leaderboard scores by identifying tasks with more reliable
performance signals, quantifying per-task score contributions, and exposing the
remaining performance gaps that are hidden by aggregate rankings.
\end{abstract}

%% file: sections/01_introduction.tex
\section{Introduction}
Improving software performance is a practical and commercially important
software-engineering task. Given a codebase and a workload, a coding agent is
expected to produce a patch that preserves correctness while improving the
program's performance
\cite{yang2024sweagent,wang2025openhands,zhang2024autocoderover}.
Code-efficiency benchmarks have also grown in scope. Earlier benchmarks focus
on function-level edits \cite{shypula2024pie}, standalone programs
\cite{huang2024effibench,du2024mercury}, and function-level and file-level code
generation tasks \cite{peng2025coffe,austin2021program,li2022competition}. Recent repository-level benchmarks such
as \sweperf \cite{fan2025sweperf}, \gso \cite{shetty2025gso}, and \sweff \cite{ma2025swe} require agents to edit real projects and
evaluate those edits with executable workloads, reference patches, and
benchmark-specific scoring rules.

Such performance benchmarks differ significantly from functional repair benchmarks because they measure a non-functional property.
Functional repair benchmarks such as SWE-Bench mainly judge whether the relevant
tests pass, a binary and usually reproducible outcome
\cite{jimenez2024swebench}. Performance benchmarks must run the base program,
the reference patch, and submitted patches under workloads, then compare noisy
runtime measurements. These measurements are not fixed quantities: they fluctuate
across runs because of CPU scheduling, cache state, memory bandwidth contention,
and machine-level microarchitectural effects
\cite{georges2007statistically,mytkowicz2009wrong,kalibera2013rigorous}.
The same patch can therefore behave faster, slower, or statistically unsupported depending on where and how it is replayed, which brings instability to the benchmarks' final evaluation.

Recent repository-level benchmarks address this problem with repeated trials,
outlier filtering, statistical tests, reference patches, and workload-selection
rules.
However, a fundamental question remains: to what extent can we trust their evaluation results, even with these countermeasures in place?
We therefore audit three recent
repository-level performance-optimization benchmarks: \sweperf{}
\cite{fan2025sweperf}, \gso{} \cite{shetty2025gso}, and \sweff{}
\cite{ma2025swe}. We ask whether official reference patches remain valid under
cross-machine replay, how scoring rules shape leaderboard rankings, and whether
replay-valid tasks still expose clear performance gaps for recent top
submissions.
Notably, these public leaderboards consist of OpenHands-based submissions, each pairing the agent scaffold with one underlying foundation model \cite{liu2025llmagents,wang2024agents,chen2025evaluatingagents}.
In the remaining text of this paper, we use \emph{submission} for the evaluated agent configuration and use model names only as shorthand labels for the underlying model component \cite{chen2021evaluating,nijkamp2023codegen,roziere2024code}.

Our systematic assessment yields three main insights.
First, the validity of reference patches is less 
reliable than what may be assumed by benchmark users.
We replay every official reference patch across four Google Cloud
machine types and three rounds, while preserving each benchmark's official
workloads and performance validity rules. Most tasks remain executable, but only $39/102$
\gso{} tasks, $11/140$ \sweperf{} tasks, and $411/498$ \sweff{} tasks satisfy
their original validity rules in all cross-machine replays. \sweperf{} is
especially fragile because many reference patches produce only close-to-zero runtime changes. This means that some official tasks lose their performance signals when
replayed on different machines, even if the task remains executable.

Second, benchmark scoring rules can strongly shape submission rankings.
Among eight public submissions shared by \gso{} and \sweff{}, the official
leaderboards disagree on 9 of the 28 pairwise submission orders. Part of this disagreement
comes from the scoring rule itself: \sweff{}'s leaderboard scoring rule gives
very large weight to low-speedup tasks, so the worst ten tasks can carry
58.5--82.8\% of a submission's score weight. A simple bounded-penalty diagnostic
changes 6/8 submission ranks and flips 8/28 head-to-head comparisons. The ranking is
therefore not only a submission comparison; it also reflects the penalty budget built
into the scoring rule.

Third, because current agent systems often use multi-agent workflows rather than
a single run, we ask an optimistic task-level question: if we look across 10
public submissions for each replay-valid task, which tasks still fall
short of the reference patch? This separates tasks already covered by at least
one public submission from tasks that still expose performance-optimization gaps.
Across 450 replay-valid \gso{} and \sweff{} tasks, at least one public submission
matches or beats the reference patch on 384 tasks; all 450 have a passing public patch; and 449 have a patch that beats the base program.
The remaining 66 tasks are therefore rarely basic correctness failures. Most already have a useful
public optimization, but that optimization does not yet reach reference-patch
speed. The remaining gap is thus less about finding any working optimization
and more about reaching or exceeding the reference-level target. As public
submissions improve, future benchmarks may need stronger reference targets to
keep separating stronger submissions from weaker ones.

Together, these results imply that leaderboard scores alone are not enough to
interpret the progress of performance-optimization agents. Readers need to know
which tasks have stable reference signals, how much the scoring rule amplifies
extreme cases, and how the remaining unsolved tasks affect the submissions'
scores.

Our contributions in this paper are as follows:
\begin{itemize}[leftmargin=1em,nosep]
    \item We replay $740$ official reference patches across four cloud machine
    types and twelve machine-round combinations, showing where each
    benchmark's reference signal remains stable and where it does not.
    \item We audit leaderboard rankings with released public outputs, showing
    that scoring rules can materially change submission rankings and that
    \sweff{}'s leaderboard scoring rule is especially sensitive to a few
    low-performing tasks.
    \item We inspect task-level outputs across public submissions,
    showing that most already have passing, faster-than-base public
    patches and that the remaining gap is mainly about reaching
    reference-patch speed.
\end{itemize}

The rest of the paper is organized as follows. Section~II defines the benchmark
selection and research questions. Section~III studies cross-machine
reference-patch validity. Section~IV audits scoring-rule sensitivity.
Section~V asks how many replay-valid tasks are covered when we look across 10
public submissions for each task. Section~VI discusses
implications for benchmark users, agent builders, and future benchmark design.

%% file: sections/03_study_design.tex
\section{Study Design}
\label{sec:study-design}

\subsection{Benchmark Selection}

We study three repository-level benchmarks that directly target coding agents on
performance optimization. 
Note that we are not
surveying every efficiency benchmark for code models, nor every open-source
benchmark artifact. We include benchmarks only when they (1) evaluate
repository-level edits, (2) target runtime optimization with executable
workloads or tests that compare code states, and (3) appear as recent, visible
peer-reviewed benchmark papers rather than leaderboard-only, code-only, or
unpublished proposals. These criteria select \gso{}, \sweperf{}, and \sweff{}
because they combine executable tasks with reviewed task definitions and are
likely to be reused to evaluate claims about agents' performance-engineering
ability.

\begin{table}[!htbp]
\caption{Repository-level performance-optimization benchmarks.}
\label{tab:benchmark-provenance}
\centering
\scriptsize
\setlength{\tabcolsep}{2.5pt}
\begin{tabularx}{\columnwidth}{@{}
>{\raggedright\arraybackslash}p{0.17\columnwidth}
>{\raggedright\arraybackslash}p{0.21\columnwidth}
>{\raggedright\arraybackslash}X@{}}
\toprule
Benchmark & Scale & Runtime performance comparison tests \\
\midrule
\begin{tabular}[t]{@{}l@{}}\gso{}\\\emph{NeurIPS 2025}\end{tabular} &
102 tasks from 10 repos &
Generated performance tests selected to compare base and reference optimization
commits
\cite{shetty2025gso}. \\
\cmidrule(l{0.5em}r{0.5em}){1-3}
\begin{tabular}[t]{@{}l@{}}\sweperf{}\\\emph{ICML 2026}\end{tabular} &
140 tasks from 9 repos &
Unit tests filtered from source repositories to compare original and
PR-modified commits
\cite{fan2025sweperf}. \\
\cmidrule(l{0.5em}r{0.5em}){1-3}
\begin{tabular}[t]{@{}l@{}}\sweff{}\\\emph{ICML 2026}\end{tabular} &
498 tasks from 9 repos &
Annotated workload scripts compare pre-edit and expert-optimized commits
\cite{ma2025swe}. \\
\bottomrule
\end{tabularx}
\end{table}

\subsection{Research Questions}

\noindent\textbf{\rqone: Do official reference patches remain valid under
cross-machine replay?}

\noindent \textit{Motivation:} Code performance-optimization benchmarks are
machine-agnostic: their tasks ask agents to optimize a program, without
targeting a particular CPU model or machine configuration. This setup assumes
that each benchmark's reference patch, together with the workload used to
distinguish patches, produces a performance signal that survives reasonable
machine variation. Cross-machine replay is therefore important: if the reference
optimization is not stable across machines, later submission comparisons on the same
benchmark rest on an unstable target. The replay checks whether each task
remains evaluable, faster than base, and valid by its original benchmark rule.

\noindent\textbf{\rqtwo: How do leaderboard scoring rules shape submission rankings?}

\noindent \textit{Motivation:} A benchmark compresses hundreds of task outcomes
into one submission score. This compression decides which
submission appears stronger. In performance optimization, the same patches can look
different depending on whether the rule counts only patches that match or beat the
benchmark reference patch, gives partial credit for smaller speedups, or heavily
penalizes very slow tasks. If a submission's rank depends on
these scoring choices, then the leaderboard ranking alone does not tell readers
what kind of progress the submission made. It is therefore necessary to make the
scoring rules explicit, compare shared public submissions under them, and audit
whether rankings are driven by broad task performance or by a small set of
high-leverage scoring cases.

\noindent\textbf{\rqthree: Do benchmark tasks remain hard for top submissions?}

\noindent \textit{Motivation:} In practice, current agent systems often use
multi-agent workflows rather than relying on a single run
\cite{schick2023toolformer,wu2023autogen}. However, the public leaderboards we
analyze rank individual OpenHands-based submissions, each pairing the agent
scaffold with one underlying model \cite{wang2025openhands}. This mismatch
matters because different submissions may solve different tasks, so a
single leaderboard entry should not be treated as the boundary of task
solvability. We therefore inspect task-level outputs across public submissions
to estimate how many tasks are covered by at least one submission and
whether the remaining gaps come from correctness, speedup depth, or optimization
strategy.

%% file: sections/04_rq1.tex
\section{\rqone: Cross-Machine Reference-Patch Validity}

We replay every official reference patch on four Google Cloud machine profiles
and reapply each benchmark's original validity rule. The goal is not to invent a
stricter benchmark, but to test whether the original reference signal survives
reasonable cross-machine variation.

\subsection{Experiment Design}

\noindent\textbf{Machine selection.} To test whether official reference patches
remain faster than the unoptimized base under common machine differences, we use
four Google Cloud machine
profiles\footnote{\url{https://docs.cloud.google.com/compute/docs/general-purpose-machines}}
with the same resource configuration: 64 vCPUs and 256GB of memory. This
resource configuration also matches the Google Cloud \texttt{n2-standard-64}
setup reported by \gso{} and \sweff{} in their original benchmark experiments
\cite{shetty2025gso,ma2025swe}. We therefore anchor the replay campaign in the
same cloud provider and resource configuration, then vary the processor platform
around that baseline. The profiles cover the two main server CPU vendors, Intel
and AMD, and include both older and newer Google Cloud processor generations.
This keeps the cloud provider and resource configuration fixed while varying
the kind of hardware difference that benchmark users are likely to encounter in
practice (Table~\ref{tab:replay-machines}).

\begin{table}[!htbp]
\centering
\scriptsize
\begin{threeparttable}
\caption{Replay machine profiles.}
\label{tab:replay-machines}
\begin{tabularx}{\columnwidth}{@{}
>{\raggedright\arraybackslash}p{0.29\columnwidth}
>{\raggedright\arraybackslash}X
>{\raggedleft\arraybackslash}p{0.17\columnwidth}@{}}
\toprule
Google Compute Engine & Processor platform &
\begin{tabular}[c]{@{}r@{}}Google Cloud\\release year\end{tabular} \\
\midrule
\texttt{n2-standard-64} &
Intel Cascade Lake, 2nd Gen Xeon, 2.80GHz &
2019 \\
\cmidrule(l{0.5em}r{0.5em}){1-3}
\texttt{n2d-standard-64} &
AMD Milan, 3rd Gen EPYC, 7B13 &
2021 \\
\cmidrule(l{0.5em}r{0.5em}){1-3}
\texttt{n4-standard-64} &
Intel Emerald Rapids, 5th Gen Xeon, Platinum 8581C &
2024 \\
\cmidrule(l{0.5em}r{0.5em}){1-3}
\texttt{n4d-standard-64} &
AMD Turin, 5th Gen EPYC, 9B45 &
2025 \\
\bottomrule
\end{tabularx}
\vspace{0.35em}
\begin{tablenotes}
\scriptsize
\item[] \emph{Note:} Processor labels combine Google Cloud platform names
with checked-in \texttt{lscpu} metadata; release years identify replay
diversity over Google Cloud generations.
\end{tablenotes}
\end{threeparttable}
\end{table}

\noindent\textbf{Experiment setup.} We follow each benchmark's guidance and
reuse the official tasks, reference patches, selected tests, and workload
definitions.
For each benchmark, we use the latest public repository version available
before our April 30, 2026 data-collection cutoff.


\noindent\textbf{Evaluation metrics.} We report two task-level metrics because
reference stability has two layers. \emph{Faster-than-base} checks the minimum
performance direction: the reference patch must beat the base program in every
replay round and machine,
$T_{\mathit{ref},r,m} < T_{\mathit{base},r,m}$. \emph{Original-rule valid}
checks the benchmark's own construction standard: the replay must still satisfy
the rule that originally qualified the task as a performance-optimization
instance. Table~\ref{tab:rq1-construction-rules} lists those original rules
from the benchmark papers \cite{shetty2025gso,fan2025sweperf,ma2025swe}.

\begin{table}[!htbp]
\centering
\scriptsize
\caption{Original benchmark validity rules.}
\label{tab:rq1-construction-rules}
\begin{tabularx}{\columnwidth}{@{}
>{\raggedright\arraybackslash}p{0.18\columnwidth}
>{\raggedright\arraybackslash}X@{}}
\toprule
Benchmark & Original validity rule \\
\midrule
\gso{} &
Each selected generated test must still pass correctness/equivalence checks and
keep a replay-time speedup of at least $1.2\times$ over the base commit. For
final low-test tasks, tests with speedup above $1.1\times$ are treated as
fallback-compatible, matching GSO's low-test construction fallback. \\
\cmidrule(l{0.5em}r{0.5em}){1-2}
\sweperf{} &
Each selected efficiency unit test must pass on the original and modified
commits. After 20 timing repetitions and IQR outlier filtering, the recomputed
Mann--Whitney minimum-gain value must remain above the dataset threshold:
$\delta_i > 0.05$. \\
\cmidrule(l{0.5em}r{0.5em}){1-2}
\sweff{} &
The official workload and correctness guards must run successfully. The base
mean workload runtime minus the expert-patched mean workload runtime must be
larger than twice the post-edit workload-runtime standard deviation. \\
\bottomrule
\end{tabularx}
\end{table}

\begin{figure}[!bp]
\centering
\includegraphics[width=\linewidth]{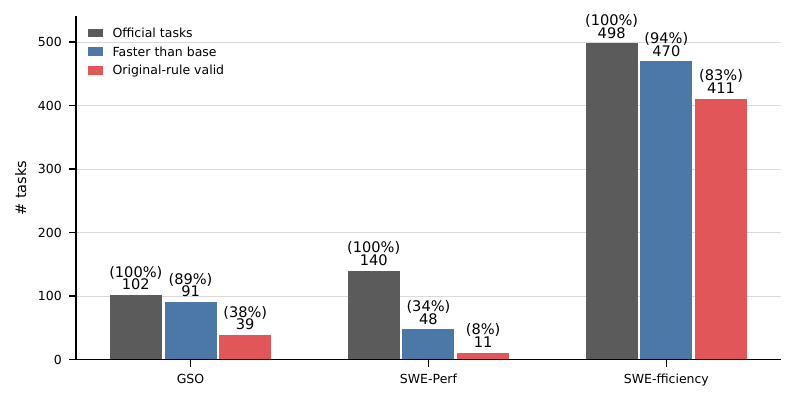}
\caption{Task counts after each replay check. Counts require passing all
12 machine-round replays (four machines across three rounds).}
\label{fig:rq1-funnel}
\end{figure}

\begin{figure*}[!t]
\centering
\includegraphics[width=\textwidth]{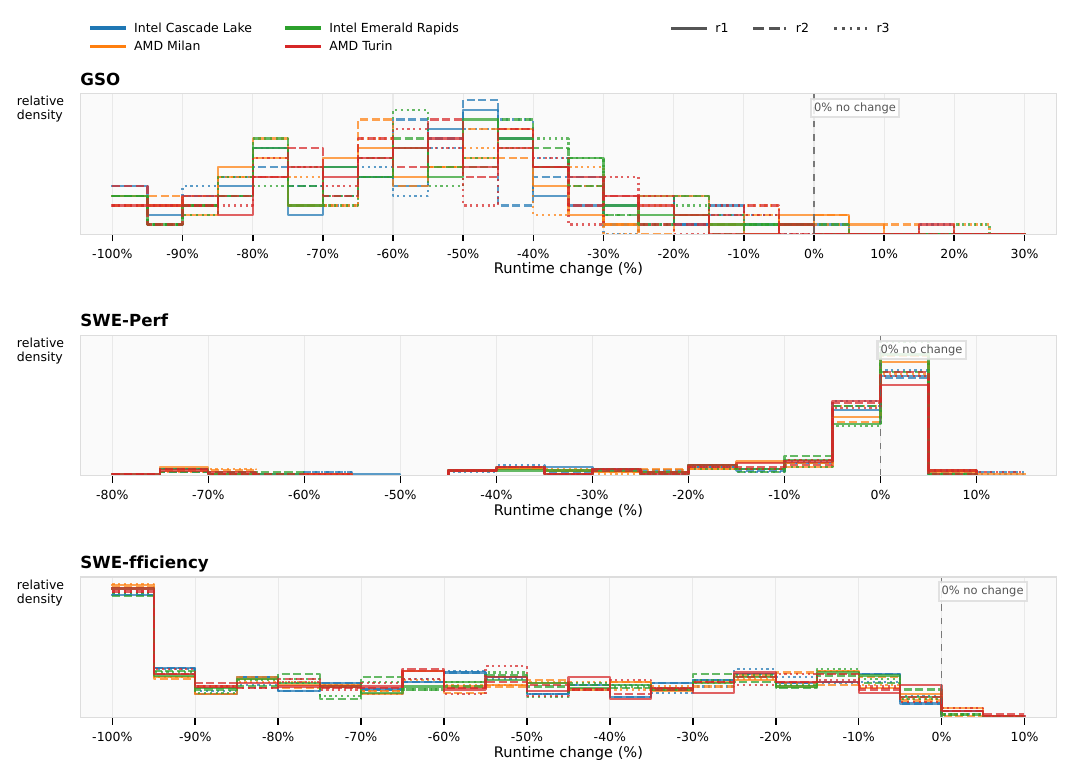}
\caption{Reference-patch runtime-change distributions by machine and replay
round. Negative values mean the reference patch reduces runtime; positive values
mean it increases runtime. \sweperf{} concentrates near the zero-change
boundary, while \gso{} and \sweff{} show larger runtime reductions.}
\label{fig:rq1-runtime-overlay}
\end{figure*}

\subsection{RQ1.1 Replay Evaluability}

\textbf{Results.}
We first check whether each reference patch replay completes and returns a usable
score. Table~\ref{tab:rq1-score-availability} counts tasks that complete
in all three rounds on each replay machine. Most tasks are fully evaluable:
\gso{} has 4/102 tasks that are not evaluable, \sweperf{} has 2/140, and
\sweff{} has 0/498.

\begin{table}[!htbp]
\centering
\scriptsize
\caption{Replay evaluability. Complete-by-machine counts require all three
rounds on n2/n2d/n4/n4d to produce usable scores.}
\label{tab:rq1-score-availability}
\begin{tabular}{@{}lccc@{}}
\toprule
Benchmark & Official & Complete by machine & Not evaluable \\
\midrule
\gso{} & 102 & 99 / 98 / 99 / 99 & 4 \\

\sweperf{} & 140 & 138 / 138 / 138 / 138 & 2 \\

\sweff{} & 498 & 498 / 498 / 498 / 498 & 0 \\
\bottomrule
\end{tabular}
\end{table}

\gso{} failures come from one functional-equivalence failure, one missing
external-data dependency,
one image-decoding failure, and one x86\_64 CPU-instruction portability failure.
\sweperf{} failures come from missing dependencies in the evaluation image.
\sweff{} has no replay-evaluability failure under our replay setup. We reported
all four \gso{} cases and both \sweperf{} cases to the benchmark maintainers.

\findingbox{Finding RQ1.1}{Most reference optimizations are replayable, but a
small number fail because external resources, build/evaluation dependencies, or
CPU-specific instructions do not port across machines.}

\subsection{RQ1.2 Reference Validity Under Replay}

We next apply the two metrics from the experiment design. The
\emph{faster-than-base} metric only asks whether the reference patch beats the
base commit in every machine-round replay. The \emph{original-rule valid}
metric is stricter: it reapplies the benchmark's own task-inclusion rule.
Figure~\ref{fig:rq1-funnel} summarizes both results using each benchmark's
official task count as the denominator; not-evaluable tasks remain in that
denominator and are counted as not passing the later timing checks.

\textbf{Results.}
Figure~\ref{fig:rq1-funnel} shows the deficiencies of reference patches. 
A simple runtime check shows that
the reference patches in only 91/102 \gso{} tasks, 48/140 \sweperf{} tasks, and 470/498 \sweff{}
tasks ran faster than the base code across machines. And, reapplying each benchmark's original validity rule leaves only 39, 11, and 411 valid
tasks. The drop is the largest for \sweperf{}, where 129 reference patches no
longer show statistically supported gain across machines.

\findingbox{Finding RQ1.2}{All three benchmarks have reference patches that run
slower than the base program in at least one cross-machine replay; applying each
benchmark's own construction rule leaves even fewer tasks valid.}

\subsection{RQ1.3 Runtime Signal and Replay Variation}

We next examine why tasks fail the original-rule check. Because the benchmarks
report performance in different units, we convert every usable replay to a
common \emph{runtime change percentage}. Negative values mean the reference
patch runs faster than the base program; positive values mean it runs slower.
For \gso{} and \sweff{}, replay outputs are speedup ratios: a speedup of $s$
means the reference patch runs $s$ times faster than the base program
($s=T_{\mathit{base}}/T_{\mathit{ref}}$). We convert them to runtime change as
$1/s-1$; for example, $1.20\times$ speedup becomes a $-16.7\%$ runtime change. For \sweperf{},
the benchmark already reports a significance-aware runtime-reduction score; we
only flip its sign so that faster reference patches are negative on the shared
runtime-change scale.

Table~\ref{tab:rq1-runtime-signal} reports both signal size and replay
variation
\cite{georges2007statistically,mytkowicz2009wrong,kalibera2013rigorous}. The
median runtime change shows how far the reference patch is from
no change, while the within-task standard deviation shows how much the 12
machine-round replays vary. We also report
$\mathrm{std./signal}=\mathrm{median\ std.}/|\mathrm{median\ change}|$; larger
values mean replay variation is large relative to the measured speed change.

\begin{table}[!tbp]
\caption{Runtime signal and replay variation.}
\label{tab:rq1-runtime-signal}
\centering
\scriptsize
\resizebox{\columnwidth}{!}{%
\begin{tabular}{lrrrr}
\toprule
Benchmark & Tasks & Median runtime change & Median runtime std. & Std./signal \\
\midrule
\gso{} & 98 & -54.20\% & 3.81pp & 0.07$\times$ \\

\sweperf{} & 138 & -0.03\% & 1.41pp & 43.23$\times$ \\

\sweff{} & 498 & -56.04\% & 2.41pp & 0.04$\times$ \\
\bottomrule
\end{tabular}
}
\end{table}

\textbf{Results.}
The table rules out a simple ``more machine noise'' explanation for
\sweperf{}. Its median within-task standard deviation is 1.41 percentage points,
lower than \gso{} and \sweff{}. The problem is signal size: the median
\sweperf{} runtime change is only -0.03\%, so even small timing shifts can cross
zero or lose statistical support. Figure~\ref{fig:rq1-runtime-overlay} shows
the same pattern: 101/138 \sweperf{} tasks have median changes within five
percentage points of zero. In contrast, \gso{} and \sweff{} have much larger
median runtime reductions, so similar or larger absolute replay variation rarely
changes the task-level conclusion.

\findingbox{Finding RQ1.3}{\sweperf{} replay failures are mainly a small-signal
problem: its reference patches cluster near no runtime change, while \gso{} and
\sweff{} reference patches usually have larger speedup margins.}

\subsection{\rqone{} Summary}

The cross-machine replay gives a three-step answer. First, most reference
patches are runnable across machines, so the main issue is not basic
evaluability. Second, runnable does not imply benchmark-valid: all three
benchmarks contain reference patches that become slower than the base program in
at least one machine-round, and reapplying each benchmark's original
construction rule narrows the valid set to 39 \gso{} tasks, 11 \sweperf{} tasks,
and 411 \sweff{} tasks. Third, the failures are not explained simply by larger
runtime noise. \sweperf{} reference patches cluster near zero runtime change,
so small timing shifts can erase statistical support; \gso{} and \sweff{}
usually retain much larger runtime reductions. One likely reason is benchmark
construction: \sweperf{} derives efficiency checks from existing repository
unit tests and accepts reference gains above a 5\% threshold, whereas \gso{} and
\sweff{} rely on more explicitly performance-oriented workloads or generated
stress tests. Future performance-optimization benchmarks should validate that
their workloads stress the optimized path strongly enough and that accepted
reference patches have clear margins over runtime noise; otherwise,
machine-level noise can be mistaken for a patch-level performance difference.

%% file: sections/05_rq2.tex
\section{\rqtwo: How Do Benchmark Scoring Rules Shape Submission Rankings?}

This section audits how released task-level outputs become the scores in leaderboard rankings. The
goal is to separate two effects: what submitted patches did on individual tasks,
and how each benchmark scoring rule weights those outcomes into one leaderboard
score.

\subsection{Experiment Design}

\noindent\textbf{Benchmark selection.} We include benchmarks that expose the
artifacts needed for a metric audit: public submissions,
per-task evaluation outcomes, and the final score used to rank submissions. As of our April 30,
2026 data-collection cutoff, \gso{} and \sweff{} released such public ranking
data. \sweperf{} did not have comparable public agent-output data,
so it is excluded from this analysis.

\noindent\textbf{Scoring rules.} To interpret rank shifts, the scoring rule must
be stated at the task level: what does one task contribute to a submission's final
score? The answer differs sharply across the two benchmarks.

\noindent\textbf{\gso{} (binary reference-level gate).} A submitted patch is
counted as either a success or a failure \cite{shetty2025gso}. For a submission
$m$ over $N$ tasks, the submission score is
\[
\mathrm{OPT@1}(m)=100\cdot\frac{\#\text{reference-level successes}}{N}.
\]
A reference-level success is a correct submitted patch whose speedup matches or
exceeds the official reference patch.
\begin{itemize}
\item \textbf{Reward.} A correct reference-level patch adds
one success, worth at most $100/N$ score points. For GSO's 102 tasks, one more
success is worth 0.98 points.
\item \textbf{Penalty.} An incorrect or below-reference
patch adds zero successes; 
it loses $100/N$
points. Correct but below-reference speedups receive no partial credit.
\end{itemize}

\noindent\textbf{\sweff{} (SpeedUp Ratio with harmonic mean).} Instead of a
binary gate, \sweff{} compares the submitted patch's speedup with the reference
patch's speedup \cite{ma2025swe}. Its task score is the SpeedUp Ratio (SR):
\[
\mathrm{SR}_{m,i}=\frac{\mathrm{speedup}_{m,i}}{\mathrm{speedup}_{ref,i}} .
\]
A value below 1 means the submitted patch is slower than the reference patch; a value
above 1 means it beats the reference patch on that workload. The released
task outputs treat failed or incorrect patches as no effective edit, which can place
them far below the reference patch. The benchmark aggregates SR values with a
harmonic mean \cite{bullen2003means,de2016harmonic} after flooring each SR at
0.001:
\[
\mathrm{HM}(m)=
\frac{N}{\sum_i 1/\max(\mathrm{SR}_{m,i},0.001)} .
\]
Here, each task contributes $1/t_i$ to the denominator, where
$t_i=\max(\mathrm{SR}_{m,i},0.001)$. Lower denominator means higher final score.

\begin{itemize}
\item \textbf{Reward.} A task that matches the reference patch has $SR=1$ and
contributes one denominator unit. If it beats the reference patch with $SR=s>1$,
it contributes only $1/s$, so the reward is
$1-1/s$ denominator units. For example, $SR=2$ saves 0.5 units and $SR=10$
saves 0.9 units; the reward is capped below one unit.
\item \textbf{Penalty.} A below-reference task exceeds the
$SR=1$ baseline by $1/SR-1$ denominator units: $SR=0.5$ adds 1, $SR=0.01$ adds
99, and the official floor $f=0.001$ adds 999. In the 498-task \sweff{}
evaluation, one floor-level task can outweigh a full reference-matching
denominator.
\end{itemize}

\subsection{RQ2.1 Cross-Benchmark Metric Sensitivity}

We first ask whether the shared public submissions rank similarly across the two
benchmarks, and whether the strong/weak judgments change when the same task
outputs are scored by another benchmark's rule. The comparison uses the eight
public submissions that appear on both \gso{} and \sweff{}. For each benchmark's
released per-task outputs, we keep the outputs fixed and change only the
aggregation rule: \sweff{} outputs are rescored with the \gso{} reference-level
gate, and available \gso{} outputs are rescored with \sweff{}'s harmonic rule.

\textbf{Results.}
The official ranks are not stable across benchmarks. The eight shared
submissions form $\binom{8}{2}=28$ unique head-to-head pairs when each pair is
compared once; the two official leaderboards disagree on 9 of them, and the
GPT-5-labeled submission moves by five positions
(Table~\ref{tab:rq2-shared-ranks}).

\begin{table}[!h]
\caption{Official ranks for shared submissions.}
\label{tab:rq2-shared-ranks}
\centering
\scriptsize
\begin{threeparttable}
\begingroup
\setlength{\tabcolsep}{3pt}
\begin{tabular}{@{}lrrrrc@{}}
\toprule
Model label & \multicolumn{2}{c}{\gso{}} & \multicolumn{2}{c}{\sweff{}} & Move \\
\cmidrule(lr){2-3}\cmidrule(lr){4-5}
& Rank & Score & Rank & Score & \\
\midrule
Claude Opus 4.6 & 1 & 41.18 & 3 & 0.1553 & $\downarrow$2 \\

GPT-5.2 & 2 & 27.45 & 4 & 0.1482 & $\downarrow$2 \\

Claude Opus 4.5 & 3 & 26.47 & 1 & 0.2250 & $\uparrow$2 \\

Gemini 3 Pro & 4 & 18.63 & 7 & 0.1024 & $\downarrow$3 \\

Claude Sonnet 4.5 & 5 & 14.71 & 5 & 0.1162 & -- \\

Gemini 3 Flash & 6 & 9.80 & 6 & 0.1056 & -- \\

GPT-5 & 7 & 6.86 & 2 & 0.1571 & $\uparrow$5 \\

Gemini 2.5 Pro & 8 & 3.92 & 8 & 0.0313 & -- \\
\bottomrule
\end{tabular}
\endgroup
\vspace{0.2em}
\begin{tablenotes}
\scriptsize
\item[] \emph{Note.} Move is \gso{} rank $\to$ \sweff{} rank.
\end{tablenotes}
\end{threeparttable}
\end{table}

Changing only the scoring rule explains part, but not all, of the mismatch.
Re-aggregating the same \sweff{} task outputs with a \gso{}-style
reference-level pass rate
raises Spearman rank correlation with \gso{} from 0.452 to 0.762 and reduces
discordant pairs from 9 to 6. In the other direction, applying \sweff{}'s
harmonic scoring to available \gso{} per-task submission reports gives a Spearman
rank correlation of 0.238 against the official \sweff{} ranking and flips 11/28
head-to-head pairs. This second rescoring makes agreement weaker than the
official comparison and flips more pairs, which suggests that the harmonic
scoring rule itself is a sensitive part of the benchmark. We therefore treat
these rescoring results as diagnostics rather than replacement rankings. They
show that scoring rules affect the ordering, while the task set and per-task
outputs also matter
(Table~\ref{tab:rq2-scoring-diagnostics}).

\begin{table}[!h]
\caption{Scoring-rule diagnostics for the eight shared submissions.}
\label{tab:rq2-scoring-diagnostics}
\centering
\scriptsize
\resizebox{\columnwidth}{!}{%
\begin{tabular}{lrrl}
\toprule
Comparison & Spearman corr. & Pair flips & Rank movement \\
\midrule
Official \gso{} vs. official \sweff{} & 0.452 & 9/28 & max 5 \\

Both with \gso{} scoring & 0.762 & 6/28 & max 3 \\

Both with \sweff{} scoring & 0.238 & 11/28 & max 5 \\
\bottomrule
\end{tabular}
}
\par\vspace{0.45em}
\begin{minipage}{0.96\columnwidth}
\scriptsize \emph{Note.} Spearman corr. compares the two rank orderings
(-1 reverse, 1 identical) \cite{spearman1904association,kendall1938rank}; pair
flips count disagreements among the 28 unique head-to-head pairs.
\end{minipage}
\end{table}

\findingbox{Finding RQ2.1}{The same submissions can rank differently across
performance benchmarks because scoring rules decide which task outcomes count
and how strongly they shape the final order.}

\subsection{RQ2.2 Low-Speedup Tail Dominance}

The scoring-rule comparison above suggests that \sweff{}'s harmonic scoring rule
deserves a closer audit. The concern is not simply that low-SR tasks receive
low scores. In a harmonic mean, each task enters the denominator as
$1/\max(SR,0.001)$, so a very low-SR task can have much more leverage than a
typical task. To measure this leverage directly, we compute each task's share of
the score denominator,
\[
w_i=\frac{1/\max(SR_i,0.001)}{\sum_j 1/\max(SR_j,0.001)} ,
\]
then sort tasks by $w_i$ within each shared submission and ask how much score weight
is carried by the worst 1, 5, and 10 tasks.

\begin{figure}[!h]
\centering
\includegraphics[width=\linewidth]{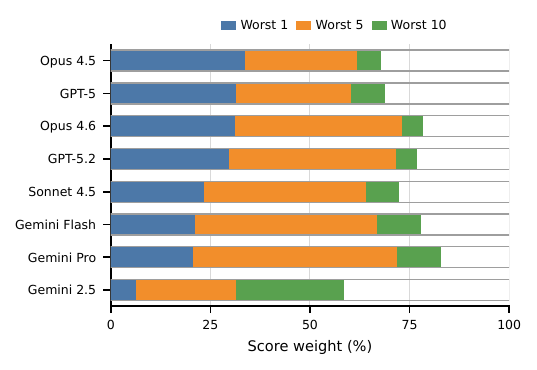}
\caption{Score weight of the worst 1, 5, and 10 tasks under \sweff{}'s
harmonic mean.}
\label{fig:rq2-burden}
\end{figure}

\textbf{Results.}
Figure~\ref{fig:rq2-burden} shows 
that Worst-1 tasks
carry 6.3--33.6\%, worst-5 tasks carry 31.4--73.1\%, and worst-10 tasks carry
58.5--82.8\% of the official score denominator. This is large for a 498-task
benchmark: a small set of low-SR tasks can explain most of the aggregate score
weight. For example, one near-floor task in Claude Opus 4.5 has raw SR 0.00134
and carries 33.6\% of the submission's score weight.

This does not mean severe slowdowns should be ignored. Penalizing them is a
reasonable benchmark choice. The issue is interpretability: when 10 tasks carry
more than half of the denominator, the official harmonic mean behaves partly as
a metric driven by the worst tasks. A rank difference may therefore reflect broad performance
across the benchmark, or it may mostly reflect a few near-floor failures. This
distinction matters when interpreting the ranking.

\findingbox{Finding RQ2.2}{\sweff{}'s final score is dominated by a few very
low-speedup tasks: the worst ten tasks carry 58.5--82.8\% of score weight, and one
near-floor task can carry one third of a submission's score weight.}

\subsection{RQ2.3 Exploring Reasonable Penalty Bounds}

After measuring tail dominance, we ask how large the penalty for one slow or
invalid task should be. The official \sweff{} floor $f=0.001$ allows one
near-zero SpeedUp Ratio (SR) task to contribute 1000 denominator units, or 999
more than an $SR=1$ reference-matching task. In a 498-task benchmark, this means
one severe failure can dominate many ordinary tasks, making the leaderboard score
hard to interpret. We therefore explore a simple bounded-penalty diagnostic: keep harmonic
aggregation, but make the largest below-reference penalty comparable to the
largest possible above-reference reward. Because the harmonic reward is capped
below one denominator unit, we cap the extra penalty at one unit as well. This
corresponds to raising the floor to $0.5$: a very low-SR task contributes at most
two denominator units, only one more than a neutral task. This probes how
rankings change when slow tasks are penalized but cannot dominate.

\begin{table}[!h]
\caption{Bounded-penalty diagnostic with floor $f=0.5$.}
\label{tab:rq2-balanced-hm}
\centering
\scriptsize
\begin{threeparttable}
\begingroup
\setlength{\tabcolsep}{2pt}
\begin{tabularx}{\columnwidth}{@{}r>{\raggedright\arraybackslash}Xrrrrr@{}}
\toprule
Rank & Model & Move &
Official HM & HM ($f=0.5$) & Med. SR & \# SR$>$1 \\
\midrule
1 & Claude Opus 4.6 & $\uparrow$2 & 0.1553 & 0.952 & 1.006 & 258 \\
2 & GPT-5.2 & $\uparrow$2 & 0.1482 & 0.944 & 1.004 & 259 \\
3 & Claude Opus 4.5 & $\downarrow$2 & 0.2250 & 0.904 & 0.986 & 212 \\
4 & Gemini 3 Pro & $\uparrow$3 & 0.1024 & 0.888 & 0.993 & 235 \\
5 & GPT-5 & $\downarrow$3 & 0.1571 & 0.867 & 0.950 & 228 \\
6 & Gemini 3 Flash & -- & 0.1056 & 0.818 & 0.919 & 185 \\
7 & Claude Sonnet 4.5 & $\downarrow$2 & 0.1162 & 0.767 & 0.858 & 164 \\
8 & Gemini 2.5 Pro & -- & 0.0313 & 0.613 & 0.497 & 38 \\
\bottomrule
\end{tabularx}
\endgroup
\vspace{0.15em}
\begin{tablenotes}
\scriptsize
\item[] \emph{Note.} Move is relative to the official \sweff{} rank (floor
$f=0.001$).
\end{tablenotes}
\end{threeparttable}
\end{table}

\textbf{Results.}
Table~\ref{tab:rq2-balanced-hm} shows that this design choice changes six of
eight submission ranks and flips 8/28 pairwise submission orders relative to the official
\sweff{} ranking. With the cap, ranks become closer to median SR and
above-reference task counts: Claude Opus 4.6 and GPT-5.2 move above Claude Opus
4.5, while GPT-5 drops. The point is not that $f=0.5$ is the right constant. It
is that the leaderboard order depends on the penalty budget given to one bad task,
which readers should keep in mind when interpreting the score.

\findingbox{Finding RQ2.3}{Changing the single-task penalty cap reshuffles the
\sweff{} leaderboard: 6/8 ranks move and 8/28 pairwise orders flip.
Leaderboard scores reflect both submission performance and penalty design.}

\subsection{\rqtwo{} Summary}

Taken together, the ranking results do not directly measure agent capability.
They combine what submissions did on individual tasks with how the benchmark turns
those task outcomes into one leaderboard score. Among the eight shared submissions,
the official \gso{} and \sweff{} rankings disagree on 9 of the 28 pairwise
orders. When the same task outputs are re-scored with another rule, ranks still
move, showing that both task outcomes and scoring design shape the final order.
The clearest difference is how much one bad task can hurt a submission. \gso{} gives
each task one equal success/failure vote. \sweff{} uses continuous speedup ratios,
but its harmonic mean with a $0.001$ floor can let a few near-zero tasks dominate
the final score: the worst ten tasks carry 58.5--82.8\% of score weight. When we
cap the maximum damage from one task, 6/8 ranks move and 8/28 pairwise
comparisons flip. Thus, \sweff{} is best read as a strict, tail-sensitive score:
its leaderboard reflects both submission behavior and the metric's penalty design.

%% file: sections/06_rq3.tex
\section{\rqthree: Do Benchmark Tasks Remain Hard for Top Submissions?}

In practice, current agent systems often use multi-agent workflows rather than
relying on a single run
\cite{yao2023react,yao2023tree,shinn2023reflexion,wu2023autogen,chen2024promise}. However, the
public leaderboards we analyze rank individual OpenHands-based submissions, each
pairing the agent scaffold with one underlying model \cite{wang2025openhands}.
This mismatch matters because different submissions may solve different tasks.
We therefore look across 10 public submissions for each replay-valid task and
ask whether any submission reaches the reference patch. This gives an optimistic
estimate of how many tasks are covered by at least one strong public submission;
it does not mean that one submission reaches the reference on all of them.

\subsection{RQ3.1 How Many Tasks Remain Below Reference?}

The first step checks the replay-valid task subset: 39
\gso{} tasks and 411 \sweff{} tasks. \sweperf{} is excluded because it has only
11 replay-valid tasks and no comparable released public agent-solution data. For
each task, we look across 10 public submissions and record whether at least one
submission reaches three milestones: passing tests, beating the unmodified base
program, and matching or beating the reference patch.

\begin{figure}[!tbp]
\centering
\includegraphics[width=\linewidth]{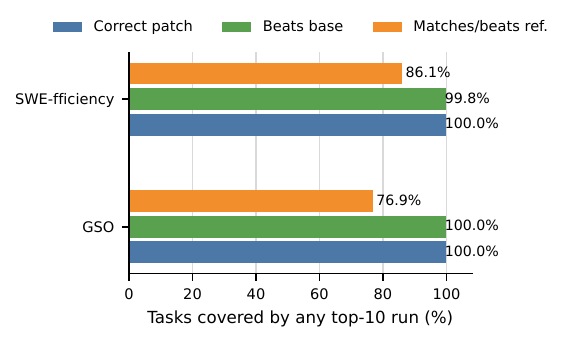}
\caption{Task outcomes across 10 public submissions on the replay-valid
subset (39 \gso{}, 411 \sweff{} tasks).}
\label{fig:rq3-layers}
\end{figure}

\textbf{Results.}
Figure~\ref{fig:rq3-layers} shows that test passing and base-program speedup are
nearly complete when we look across 10 public submissions. Every replay-valid
\gso{} and \sweff{} task has at least one passing public patch, and 449/450 tasks
have a passing patch that beats the base program. The remaining signal appears
only at the stricter reference-level gate: at least one of the 10 public
submissions matches the reference on 384/450 tasks, leaving 66 tasks not reaching
reference speed, 9/39 for \gso{} and 57/411 for \sweff{}.

\findingbox{Finding RQ3.1}{Across 10 public submissions, most replay-valid tasks
already have a strong public solution: 384/450 match or beat the reference patch,
all 450 have a passing patch, and only 66 remain below reference.}

\subsection{RQ3.2 Are the Remaining Tasks Truly Unsolved?}

The 66 tasks that still do not reach the reference-patch speed could be genuinely
unsolved, or they could already have useful public patches that are simply not
as fast as the reference patch. For each task, we inspect the \emph{best public
patch}: among the 10 public submissions, the patch with the largest
replayed speedup relative to the reference patch. Every task has a correct public patch, 65/66 already run
faster than the base program, and only one task has a best patch that passes tests without
improving runtime. Individual attempts still fail, but this task-level selection
removes most task-level blockers: across the 660 attempts behind
these 66 tasks, 160 fail functionality or validation, 3 do not complete, and 46
pass correctness without improving runtime \cite{chen2025beyond}.

\begin{figure}[!htbp]
\centering
\includegraphics[width=\linewidth]{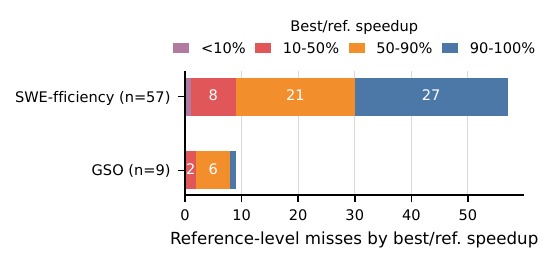}
\caption{Best public patch speed relative to the reference patch for the 66
tasks that do not reach reference-patch speed.}
\label{fig:rq3-boundary}
\end{figure}

\textbf{Results.}
Figure~\ref{fig:rq3-boundary} shows that most remaining tasks are close to the
reference patch rather than broken outright. For \gso{}, the best public patch
reaches a median 85.3\% of the reference patch's speedup and would need
1.17$\times$ more speedup to match it. For \sweff{}, the median is 87.9\%,
leaving 1.14$\times$. Many tasks are near the reference: 27 \sweff{} tasks reach
90--100\% of the reference patch's speedup. The tail still matters, however: 8
\sweff{} tasks reach only 10--50\% and one is below 10\%.

\findingbox{Finding RQ3.2}{Remaining tasks are rarely complete failures: all
have a correct public patch, 65/66 beat the base program, and the median best
patch reaches 85.3\%/87.9\% of reference speedup on the two benchmarks.}

\subsection{RQ3.3 Does the Hard Tail Require the Reference Strategy?}

The previous results show that the remaining tasks are often not basic
correctness failures: the best public patch usually passes tests and improves
runtime, but still trails the reference patch. A natural hypothesis is that
strategy choice explains this gap. If high performance depends on targeting the
right hot operation or using the right optimization mechanism, public patches
that use a different strategy from the reference patch should leave larger
speedup gaps. This subsection explores that hypothesis by comparing each best
public patch with its reference patch. The reference is a diagnostic baseline,
not an imitation target: a different strategy is acceptable if it reaches
similar speedup.

\emph{Annotation method.} For each of the 66 tasks that still do not reach
reference speed,
we compare the reference patch with the best public patch defined above. We
use the high-level optimization categories from Peng et
al.'s taxonomy~\cite{peng2026sysllmatic,peng2025agentscodeoptimization}. We reuse their automated
annotation script with GPT-5.5 as the annotation model to assign one category to
each reference patch and each best public patch. We schema-check the labels,
join them into reference-vs.-public pairs, and mark whether the best public patch
uses the same category, a different category, or no visible production
optimization. We report both category alignment and the remaining gap, measured
as best-public speedup divided by reference-patch speedup. We manually
sanity-check the labels against patch diffs.

\begin{table}[!htbp]
\caption{Reference-category alignment.}
\label{tab:rq3-reference-category-alignment}
\centering
\small
\begin{tabular}{lrrr}
\toprule
Reference category & Ref. pairs & Same cat. & Diff. cat. \\
\midrule
Algorithm & 31 & 21/31 & 10/31 \\
Structure & 16 & 7/16 & 9/16 \\
Memory & 12 & 3/12 & 9/12 \\
Build & 3 & 0/3 & 3/3 \\
Control & 3 & 1/3 & 2/3 \\
Data struct. & 1 & 0/1 & 1/1 \\
\midrule
Total & 66 & 32/66 & 34/66 \\
\bottomrule
\end{tabular}
\end{table}

\textbf{Results.}
Table~\ref{tab:rq3-reference-category-alignment} shows both the shape of the
remaining tasks and how often the best public patch matches the reference
category.
Algorithm reference patches are the largest group and are matched most often
(21/31). Structure is split (7/16 matched), while memory-heavy reference patches
are usually approached through another category (9/12 different). Overall, 32/66
best public patches use the same high-level category as the reference patch, and
34/66 use a different category or show no visible production optimization.

\begin{figure}[!htbp]
\centering
\includegraphics[width=\linewidth]{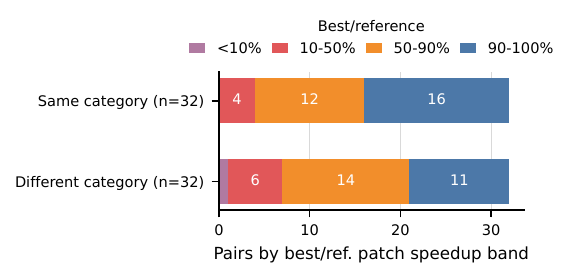}
\caption{Best public/reference speedup ranges by high-level category alignment.}
\label{fig:rq3-patch-relation-gaps}
\end{figure}

\begin{figure}[!htbp]
\centering
\includegraphics[width=0.9\linewidth]{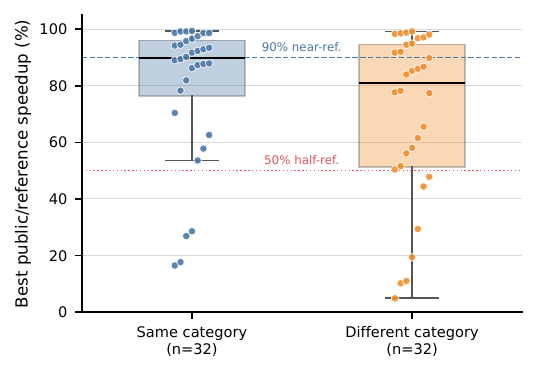}
\caption{Best public/reference speedup by high-level strategy alignment.}
\label{fig:rq3-strategy-alignment-speedup}
\end{figure}

Figures~\ref{fig:rq3-patch-relation-gaps}
and~\ref{fig:rq3-strategy-alignment-speedup} compare the remaining speed gap by
category alignment. For this speedup comparison, we exclude the two cases without
visible production optimization, leaving 32 same-category and 32
different-category public optimizations. Same-category patches are closer to the
reference on median (89.8\% vs. 81.1\%; remaining multiplier 1.12$\times$ vs.
1.23$\times$), but the two groups overlap substantially: 16/32 same-category
cases still remain below 90\% of the reference speedup, while 11/32
different-category cases reach 90--100\%. Same-category cases even include four
below-50\% gaps. Category alignment is therefore only a partial signal, not a
sufficient condition for near-reference performance. The remaining gap often
reflects implementation depth: agents often find a useful optimization idea but
do not apply it broadly or carefully enough to match the reference patch.

\findingbox{Finding RQ3.3}{Optimization strategy mismatch is not the main reason
tasks stay below reference: among 66 tasks, 32 same-category public patches
remain slower, while 11 different-category patches reach 90--100\% of reference speed.}

\subsection{\rqthree{} Summary}

Looking across 10 public submissions shows that many replay-valid \gso{}
and \sweff{} tasks are already solved by at least one public submission. Across 450
tasks, 384 already match or beat the reference patch, all 450 have a passing
public patch, and 449 have a patch that beats the base program.
The remaining 66 tasks are therefore not mostly basic correctness failures. They
usually already have a correct and runtime-improving public patch, with median
best-patch speed reaching 85.3\% of the reference speedup on \gso{} and 87.9\%
on \sweff{}.

The remaining speed gap also does not appear to be mainly a broad
strategy-selection problem. In 32/66 tasks, the best public patch uses the same
high-level optimization category as the reference patch; in 11 more, a different
category still reaches 90--100\% of the reference speedup. This suggests that
public submissions often identify a useful performance direction, not merely a
passing edit. What remains is deeper optimization after the first correct
speedup: without seeing the reference patch, the runs may stop short of the
extra iteration needed to reach reference-patch speed.

%% file: sections/07_discussion.tex
\section{Discussion}

Our results suggest three practical implications for reading and designing
performance-optimization benchmarks.

\textbf{For benchmark users: separate signal from measurement design.}
Leaderboard scores are useful, but our results show that current benchmark
scores should not be read as direct labels of agent capability.
RQ1 shows that reference-patch evidence can change across
machines, especially when the runtime gain is close to zero. This means a
reported optimization can be hard to distinguish from runtime noise. RQ2
shows a second layer of interpretation: even when task outputs are fixed, the
aggregation rule can decide how much one bad task matters. Under \sweff{}'s
leaderboard scoring rule, a few low-speedup tasks can account for most of the
score weight, so a rank difference may reflect broad task performance, a small
set of severe failures, or both. Users of these benchmarks should therefore look
beyond the leaderboard rank. At minimum, reports should distinguish
cross-machine verified tasks from unstable tasks, show per-task score weight,
and compare submissions under aggregation rules that match the use case. For some
users, strong worst-case penalties are appropriate; for others, median-speedup
or reference-level coverage is more informative.

\textbf{For agent builders: interpret multi-agent gains at the task level.}
A single submitted run remains a valid and strict competition setting, but it is
not the same as asking how many tasks today's public submission outputs still leave
unsolved. 
RQ3 shows that, on the replay-valid \gso{} and \sweff{} tasks, at least one of
the 10 public submissions already reaches the reference patch on 384/450
tasks, and almost every remaining task has a correct patch that is faster than
the base program. This matters for multi-agent workflows that can
draw on several agent configurations \cite{li2025reinforcement,ma2025agentguard}. If a new agent
configuration is evaluated only by making more tasks reach the
reference patch, the remaining room is narrow and the measured improvement may
depend heavily on a small set of tasks. This does not make the benchmarks
useless; it changes what they are good for. They are still useful for studying
the last step from ``faster than base'' to ``as fast as the reference patch,''
but less suitable as the only evidence of general performance-engineering
ability.

\textbf{For future benchmark designers: move closer to the full performance
engineering problem.}
Current benchmarks provide executable workloads and reproducible task artifacts,
but they often simplify the task by exposing the optimized code region,
workload, or stress test to the agent. Production performance work usually starts
earlier: engineers inspect profiles, flame graphs, traces, latency breakdowns,
or dashboards, then choose which hot operation in a large codebase is worth
changing, infer which measurement matters, and judge whether a local edit will
help the real workload without a ready-made stress test for immediate feedback.
Future benchmarks should preserve executable replay while adding this diagnostic
layer: hotspot localization from profiles or traces, ambiguous optimization
targets, validation against workloads not fully visible during patch search, and
resource metrics beyond one runtime number \cite{chen2026rethinking}. CPU time, latency, allocation
behavior, memory footprint, and regression risk all matter, and optimizing one
can hurt another \cite{sun2026executing}. The goal is not to discard today's
benchmarks, but to make the next ones test a harder workflow: finding the
bottleneck, choosing a justified optimization, implementing it across the right
code paths, and proving that it improves the workload that matters.

%% file: sections/08_threats.tex
\section{Threats to Validity}

\noindent\textbf{Internal validity.} Our replays and metric recomputations may
still be affected by implementation choices, hardware variation, and released
artifact quality. The replay analysis therefore uses a strict all-replay rule:
a task is replay-valid only when the reference patch satisfies the benchmark's
original construction rule in all 12 machine-round replays. This conservative
rule may undercount usable tasks, but it protects later analyses from unstable
reference signals.

\noindent\textbf{External validity.} The results cover three recent
repository-level performance-optimization benchmarks and specific public
leaderboard snapshots, not all coding-agent or performance-engineering settings.
Our analysis across 10 public submissions is also a task-coverage proxy,
not a newly engineered multi-agent workflow or a claim about any single submitted
configuration. Future underlying models, agent scaffolds, hardware, and benchmark policies may change the exact
counts.

\noindent\textbf{Data construction validity.} We rely on each benchmark's
released tasks, reference patches, public submissions, and scoring records.
\sweperf{} is excluded from the metric and RQ3 task analyses because
comparable public agent-output artifacts were unavailable.

%% file: sections/09_related_work.tex
\section{Related Work}

Code-efficiency benchmarks have expanded from small code units to full
repositories. PIE studies performance-improving edits, while EffiBench, Mercury,
ENAMEL, ECCO, EvalPerf/DPE, and ACECode evaluate whether generated or revised
code remains correct while improving efficiency
\cite{shypula2024pie,huang2024effibench,du2024mercury,qiu2025enamel,
waghjale2024ecco,liu2024evalperf,yang2024acecode,harman2025mutation}. COFFE adds function- and
file-level tracks, while EffiBench-X, KernelBench, AlgoTune, and PerfCodeBench
broaden evaluation across languages, GPU kernels, numerical programs, and
system-level high-performance code
\cite{peng2025coffe,qing2025effibenchx,ouyang2025kernelbench,
press2025algotune,jing2026perfcodebench}. At the repository and project level,
\gso{}, \sweperf{}, and \sweff{} evaluate edits to real repositories with
executable workloads, reference patches, and benchmark-specific metrics
\cite{shetty2025gso,fan2025sweperf,ma2025swe}. PerfBench, PEACE/PeacExec,
ISO-Bench, FormulaCode, and CppPerf respectively cover real-world performance
bugs, project-level efficiency optimization with hybrid editing, inference
workloads, large-codebase agent optimization, and performance-improving C++
commits
\cite{garg2026perfbench,ren2025peace,nangia2026isobench,
sehgal2026formulacode,ho2026cppperf}. Together, these benchmarks show a clear
shift from function-level efficiency to repository-scale performance
engineering
\cite{hou2024large,fan2023survey,zhang2024survey}.

In contrast, we do not introduce another benchmark or report a new agent
leaderboard. We audit the released artifacts behind existing repository-level
benchmarks: whether official reference patches replay reliably, how scoring rules
change public-submission rankings, and which replay-valid tasks still separate
public submissions from reference patches.

%% file: sections/10_conclusion.tex
\section{Conclusion}

Performance-optimization benchmarks now influence how coding agents are judged,
but their scores are not self-explanatory. Our audit of three recent
repository-level benchmarks shows why. After replaying 740 official reference
patches across four cloud machines and three rounds, only 39/102 \gso{} tasks,
11/140 \sweperf{} tasks, and 411/498 \sweff{} tasks kept their original validity
signal in every replay. Leaderboard ranks also depend on the scoring rule:
among the eight public submissions shared by \gso{} and \sweff{}, the official
rankings disagree on 9/28 pairwise orders, and the worst ten low-speedup
\sweff{} tasks carry 58.5\%--82.8\% of the score weight. Finally, many
replay-valid tasks are already covered by at least one public submission:
384/450 match or beat the reference patch, and 449/450 beat the base program.
For the remaining tasks, the problem is usually not finding any working
optimization, but closing the last speedup gap to the reference patch.

The next generation of performance-optimization benchmarks should therefore test
more than patching a pre-specified entry point under a fixed workload. They should
ask agents to reason from performance symptoms or profiles, choose optimization
targets, and validate runtime improvements together with memory, latency, and
other resource costs.

%% file: sections/11_data_availability.tex
\section*{Data Availability}

The public data are available at
\url{https://github.com/chenzhi-cz/performance-optimization-benchmark-reliability}.